\documentclass[aps,superscriptaddress,altaffilletter,lengthcheck,
tightenlines,showpacs,showkeys]{article}

\usepackage{multirow}

\newcommand{\ben}{\begin{eqnarray}}
\newcommand{\een}{\end{eqnarray}}
\newcommand{\be}{\begin{equation}}
\newcommand{\ee}{\end{equation}}
\newcommand{\ba}{\begin{eqnarray}}
\newcommand{\ea}{\end{eqnarray}}

\newcommand{\ga}{\gamma}
\newcommand{\ro}{\rho}

\usepackage[dvipdf]{epsfig}
\usepackage{color}
\begin{document}
\title{Emergent Universe as an interaction in the dark sector\\  }

\author{Emiliano Marachlian \thanks{Depto. de F\'{\i}sica, FCEyN,  Universidad de Buenos Aires and  IFIByNE-CONICET, Buenos Aires 1428, Argentina},
 Iv\'an E. S\'anchez G. \thanks{Depto. de F\'isica, FCEyN and IFIBA, Universidad de Buenos Aires, Buenos Aires, Argentina isg.cos@gmail.com and isg@df.uba.ar.}
 and Osvaldo P. Santill\'an \thanks{Depto. de Matem\'atica, FCEyN, Universidad de Buenos Aires, Buenos Aires, Argentina firenzecita@hotmail.com and osantil@dm.uba.ar.} }
\date {}
\maketitle


\begin{abstract}
A cosmological scenario where dark matter interacts with a variable vacuum energy for a  spatially flat Friedmann-Robertson-Walker space-time is proposed and analysed to show that with a linear equation of state and a particular interaction in the dark sector it is possible to get a model of an Emergent Universe. In addition, the viability of two particular models is studied by taking into account recent observations. The updated observational Hubble data and the JLA supernovae data are used in order to constraint the cosmological parameters of the models and estimate the amount of dark energy in the radiation era. It is shown that the two models fulfil the severe bounds of $\Omega_{x}(z\simeq 1100)<0.009$ at the $2\sigma$ level of Planck.

\end{abstract} 
\vskip 1cm

\section{Introduction}
Since 1998, there are strong evidences that the universe is flat and in an accelerated expansion phase. Some of these evidences comes from the cosmological and astrophysical data from type Supernovae Ia (SNIa) \cite{Riess1}, \cite{Perlm}, \cite{Astier}, the spectra of the Cosmic Microwave Background (CMB) \cite{Spergel}-\cite{Planck2015XIV} radiation anisotropies and Large Scale Structure (LSS) \cite{Bertone}, \cite{Freese}. One of the alternatives to explain this faster expansion phase is to consider a mysterious dark energy component with negative pressure. The simplest type model of dark energy corresponds to a positive cosmological constant $\Lambda$. Another important component of our Universe is dark matter, it shares the non luminous nature with the dark energy. It is gravitationally attractive and leads to the formation of large scale structures.

There are several models which attempt to explain the origin or the dynamics of the dark matter and the dark energy. Some of them propose that the origin could become form a kind of dynamical scalar field, as the quintessence model \cite{Ferreira}-\cite{Sahni}. Other models expect that the cosmological term $\Lambda$ should not be strictly constant, it appears as a smooth function of the Hubble rate $H(t)$ \cite{Grande}-\cite{Alcaniz}. It has actually been shown that, in some cases, these models can actually fit the data better than the concordance $\Lambda$CDM model at a level of $3-4\sigma$ \cite{Sola1}-\cite{Sola5}.

A considerable alternative to the $\Lambda$CDM model is the possibility of interaction in the dark sector. This non-gravitational interaction gives rise to a continuous transfer of energy between dark, energy and matter, i.e. we suppose that one component can feel the presence of the other through the gravitational expansion of the Universe \cite{Luis}. As it is expected, a connection between the dark components changes the background evolution of the dark sector \cite{Luis7}-\cite{Vali}, giving rise to a rich cosmological dynamics compared with non interacting models. It can be found  that this phenomenon seems to be possible at theoretical level when coupled scalar fields are considered \cite{Amendola1}, \cite{Amendola2} and it is also compatible with the current data coming from Planck \cite{Yang1}-\cite{Abdalla}.

As it is known, the big bang cosmology scenario has some problems both in the early and in late universe. Many of these problems emerge when one is describing the early Universe, the horizon problem, the flatness problem, fine-tuning, etc. \cite{Albrecht}, \cite{Linde}. These unresolved issues could be explained by the physics of inflation and the introduction of a small cosmological constant for late acceleration, but they are not clearly understood. An alternative is the Emergent Universe scenario, in which an inflationary universe emerges from a small static state that has within it the origin of the development of the macroscopic universe. The universe has a finite initial size and since the initial stage is Einstein static, there is no time-like singularity. 
As the Einstein static universe solution obtained is unstable, it creates fine tuning problems for emergent cosmology, it impose conditions for an appropriate choice of the inflaton potential, more precisely the initial value of the field must match the asymptotic form of the potential. Which is a consequence
of the Einstein static universe being unstable in General Relativity \cite{Bag}.
Mukherjee et al. \cite{Muk0} also showed that a successful inflation may be permitted in the Emergent Universe scenario. This model does not solve all the inflationary problems mentioned above, but because it is an ever-existing universe, there is no horizon problem \cite{Ellis}. The possibilities of an emergent universe have been studied in few papers. Del Campo \emph{et. al.} \cite{Campo} studied the emergent universe model in the context of a self-interacting Jordan-Brans-Dicke theory, Mukherjee \emph{et. al.} \cite{Muk} in the framework of general relativity, Paul and Ghose in Gauss-Bonnet gravity \cite{Paul}, in a Horava gravity was studied by Mukherjee and Chakraborty \cite{Muk1}, etc.  

The emergent universes proposed by Mukherjee \cite{Muk} are late-time de Sitter with an equation of state of the form $p=\mathcal{A}\rho-\mathcal{B}\rho^{1/2}$, where $\mathcal{A}$ and $\mathcal{B}$ are constants. This is a special case of the Chaplygin gas \cite{Bento}-\cite{Bilic}. Lately, the onset of the recent accelerating phase had been determined by the constraints of the parameters $\mathcal{A}$, $\mathcal{B}$ with the observational data \cite{Paul1}, \cite{Ghose}.

The aim of this work is two-folded. On the one hand it is shown that, by assuming the existence of an interacting dark sector with a barotropic equation of state in the context of General Relativity, an emergent universe dynamics such as the ones considered in \cite{Muk}, \cite{Paul1}, \cite{Paul2} may arise. It should be emphasized that none of the previous emergent scenarios was obtained by taking into account a linear barotropic equation of state in General Relativity. This fact is one novel feature of the present work. On the other hand, we are considering models described by a source equation which is of second order \cite{Luis} differ from those studies \cite{Muk}, \cite{Paul1} that consider a conservation equation which is of first order. Certain explicit solutions of this new models are reported in the text. In addition the explicit form of these solutions, the updated Hubble data, the JLA supernovae data points and the severe bounds reported by the Planck mission on early dark energy are used below in order to constraint the parameters of our model.

\section{Interaction Model}
In the Interaction Scenario a spatially flat isotropic and homogeneous universe described by Friedmann-Robertson-Walker (FRW) spacetime is usually considered. The universe is filled with three components, baryonic matter, and two fluids that interacts in the dark sector. The first is a decoupled component. The evolution of the FRW universe is governed by the Friedmann and conservation equations,
\be
\label{E1a}
3H^{2}=\ro_T=\rho_{r}+\rho_{b}+\ro_{m}+\ro_{x},
\ee
\be
\label{E1c}
\dot\ro_{b}+3H\ga_{b}\ro_{b}=0,
\ee
\be
\label{E1b}
\dot{\ro}_{m}+\dot{\ro}_{x}+3H(\ga_m\rho_{m}+\ga_x\rho_{x})=0,
\ee
where $H=\dot a/a$ is the Hubble expansion rate and $a(t)$ is the scale factor. The equation of state for each species, with energy densities $\ro_{\rm i}$, and pressures $p_{\rm i}$, take a barotropic form $p_{\rm i}=(\gamma_{\rm i}-1)\ro_{\rm i}$, and the constants $\ga_{\rm i}$ indicate the barotropic index of each component being ${\rm i}=\{x,m,b\}$, so  that  $\gamma_{x}=0$, $\ga_{b}=1$, whereas $\ga_{m}$ will be estimated later on. Then $\rho_{x}$ plays the role of a variable cosmological constant, $\ro_{b}$ represents a pressureless barionic matter, and $\rho_{m}$ can be associated with dark matter.


In order to continue the analysis of the interacting dark sector we note that, by separating the conservation equation for the system Eq. \ref{E1b} and by using the variable $\eta$ defined above, the following energy transfer equation between the two fluids is obtained
\be
\label{rod3}
{\ro'}_{m}+\ga_m\rho_{m}=-Q,  \qquad {\ro'}_{x}+\ga_x\rho_{x}=Q,
\ee
where the variable $\eta=ln(a/a_0)^3$ has been introduced, with $a_0$ the present value of the scale factor ($a_0=1$), and $Q$ indicates the energy exchange between the dark components. 

In the following it is assumed that there is no interaction between the baryons and the dark sector, so the energy density is conserved and the prime indicates differentiation with respect to the new time variable $'\equiv d/d\eta$. Under this situation, Eqs. (\ref{E1c}) leads to the energy density for baryonic matter, $\ro_{b}\sim a^{-3}$.


In this work we present a phenomenological interaction $Q$ between both dark components with a scale factor power law dependence as in reference \cite{Richarte}
\begin{equation}
\label{Q}
Q=-2A\sqrt{B}a^{-3r}-Ba^{-6r},
\end{equation}
where $A$ and $B$ are the coupling constants that measure the strength of the interaction in the dark sector. In this case, we will analyse the models with $r=1/2$ and $r=1/3$. These kind of interactions are now studied under the view of the new observations and gives rise to a dark energy model that can be viewed as an emergent universe \cite{Paul1}, \cite{Ghose}. 

By replacing the specific form of $Q$ into the source equation (\ref{rod3}) and the value $\ga_{x}=0$, the first order differential equations for the dark matter density $\ro_m(a)$ and the dark energy density $\rho_x(a)$ can be solved. The relation between the energy density and the redshift $z$ may be found by considering the expression of the scalar factor in terms of the redshift, $z+1=1/a$, so the solutions are
\be
\label{rom}
\ro_{m}=C_2(1+z)^{3\ga_m}+\frac{2A\sqrt{B}}{\ga_m-r}(1+z)^{3r}+\frac{B}{\ga_m-2r}(1+z)^{6r},
\ee
\be
\label{rox}
\ro_x=C_1+\frac{2A\sqrt{B}}{r}(1+z)^{3r}+\frac{B}{2r}(1+z)^{6r}.
\ee
where $C_1$ and $C_2$ are the integration constants. By adding Eqs. (\ref{rom}) and (\ref{rox}), we get the energy density of the dark sector
\[\ro=C_1+C_2(1+z)^{3\ga_m}+\frac{2A\sqrt{B}}{r}\frac{\gamma_m}{\ga_m-r}(1+z)^{3r}\]
\be
\label{rod5}
+\frac{B}{2r}\frac{\gamma_m}{\ga_m-2r}(1+z)^{6r}.
\ee
 
For the choice $C_2=0$, the energy density of Eq. (\ref{rod5}) may be written in the form  $\ro(a)=(\beta+\alpha a^{3r})^2/a^{6r}$, with $\beta$ and $\alpha$  simple constants. This form of the energy density is the one obtained in \cite{Muk}, these authors found it by using a polytropic equation of state of the form $p=\mathcal{A}\rho-\mathcal{B}\rho^{1/2}$. One purpose of this research is to show that in a spatially flat universe with a linear barotropic equation of state (instead of polytropic) and the interaction Eq. (\ref{Q}) may cast the energy density of the an Emergent Universe. That is why we studied this specific interaction and no others. Probably there are other interactions that could cast this particular energy density, maybe there is one where it could be possible to leave the barotropic index $\gamma_x$ free. We are working on this and we leave it for a future research. For the purposes of the present work, it can be seen that the models studied in the references \cite{Paul1}, \cite{Ghose} are a particular case of this model when the constant $C_2=0$. We will not analyse these models, instead, we studied the model with $C_2\neq0$ for $r=1/2$ and $r=1/3$ and we have found the constraints to determine the model parameters, for every case respectively. To the best of our knowledge, this was not done in any previous work.

By taking into account the present-density parameters $\Omega_{i0}=\ro_{i0}/3H_{0}^2$ along with the flatness condition $1=\Omega_{b0}+\Omega_{x0}+\Omega_{m0}$, the integration constants $C_1$ and $C_2$ may be expressed in terms of the observational density parameters
\be
\label{C1}
C_1=3{H_0}^2\Omega_{x0}-\frac{2A\sqrt{B}}{r}-\frac{B}{2r},
\ee
\be
\label{C2}
C_2=3{H_0}^2\Omega_{m0}-\frac{2A\sqrt{B}}{\ga_m -r}-\frac{B}{\ga_m-2r}.
\ee
In this case the Friedmann equation (\ref{E1a})  is given in terms of the redshift and density parameters by
\[3H^2(z)=(1-\Omega_{x0}-\Omega_{m0})(1+z)^{3}+C_1+C_2(1+z)^{3\ga_m}\]
\be
\label{H2}
+\frac{2A\sqrt{B}}{r}\frac{\gamma_m}{\ga_m-r}(1+z)^{3r}+\frac{B}{2r}\frac{\gamma_m}{\ga_m-2r}(1+z)^{6r}.
\ee
The specific models with $r=1/2$ and $r=1/3$ have six independent parameters ($H_0$, $\Omega_{x0}$, $\Omega_{m0}$, $A$, $B$, $\ga_{m}$) to be completely specified. The above function (\ref{H2}) will be used in the next section for analysis with observational results and to determine the model parameters. For both models, in the limit case $z\rightarrow-1$, the energy density goes to a constant value like the $\Lambda$CDM model, so the universe exhibits a de Sitter phase at late times. In the dark energy domains, the energy density Eq. (\ref{rod5}) for the model with $r=1/3$ corresponds to a cosmic fluid that behaves as a composition of cosmological constant, domain walls and cosmic strings \cite{Muk}.   

\section{Constrains on the parameters of the model}

\subsection{Observational Hubble data}
A set of measurements for Hubble parameter $H(z)$ at different redshifts \cite{Stern} \cite{Verde} \cite{Moresco} \cite{Busca} \cite{Zhang} \cite{Blake} \cite{Chuang} will be considered in the following. A qualitative estimation of the cosmological parameters for the models with $r=1/2$ and $r=1/3$ described above is found. The values of the function $H(z)$ are directly obtained from the cosmological observations, so this function plays a fundamental role in understanding the properties of the dark sector. The bibliography \cite{W-7}, \cite{WMAP9}, \cite{Ratra} shows $H_{obs}$ for different redshifts with the corresponding 1$\sigma$ uncertainties. The probability distribution for the $\theta$-parameters, for each model, is $P(\theta)=\aleph\exp^{-\chi^2(\theta)/2}$ \cite{Press}, being $\aleph$ a normalization constant. In order to obtain the parameters of the models we first minimized a chi-square function $\chi^2$ defined as 
\be
\label{chi}
\chi^2(\theta)=\sum^{N=29}_{i=1}\frac{[H(\theta;z_i)-H_{obs}(z_i)]^2}{\sigma^2(z_i)},
\ee
where $H_{obs}(z_i)$ and $H(\theta,z_i)$ are the observed and observational values of the Hubble parameter $H(z)$ at different redshifts $z_i$ and $\sigma(z_i)$ is the corresponding 1$\sigma$ error. The Hubble function $H(\theta,z_i)$ is (\ref{H2}) evaluated at $z_i$, for both models, with $r=1/2$ and $r=1/3$ respectively. The variable $\chi^2$ is a random variable that depends on $N=29$, the number of the data, and its probability distribution is a $\chi^2$ distribution for $N-n$ degrees of freedom, with $n=2$, where $n$ is the number of parameters. The $\chi^2$ function reaches its minimum value at the best fit value $\theta_c$ and the fit is good when $\chi^{2}_{min}(\theta_c)/(N-n)$ is close to one \cite{Press}. For a given pair $(\theta_1,\theta_2)$ of independent parameters, fixing the other ones, the confidence levels (C.L.) $1\sigma$ $(68.3\%)$ or $2\sigma$ $(95.4\%)$ will satisfy $\chi^2(\theta)-\chi^{2}_{min}(\theta_c)\leq2.30$ or $\chi^2(\theta)-\chi^{2}_{min}(\theta_c)\leq6.17$ respectively.

\begin{figure}
\centering
\resizebox{1.8in}{!}{\includegraphics{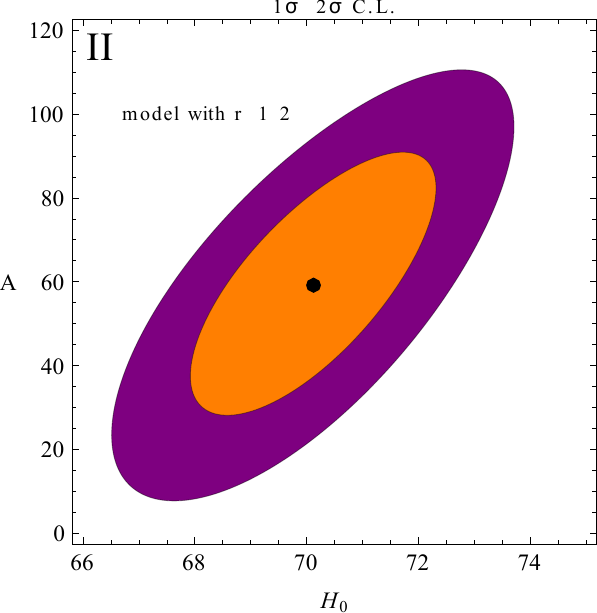}}\hskip0.06cm
\resizebox{1.8in}{!}{\includegraphics{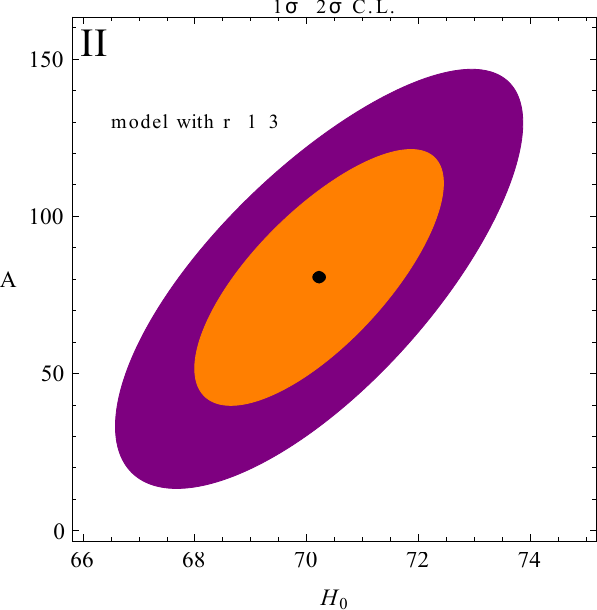}}\hskip0.06cm
\resizebox{1.8in}{!}{\includegraphics{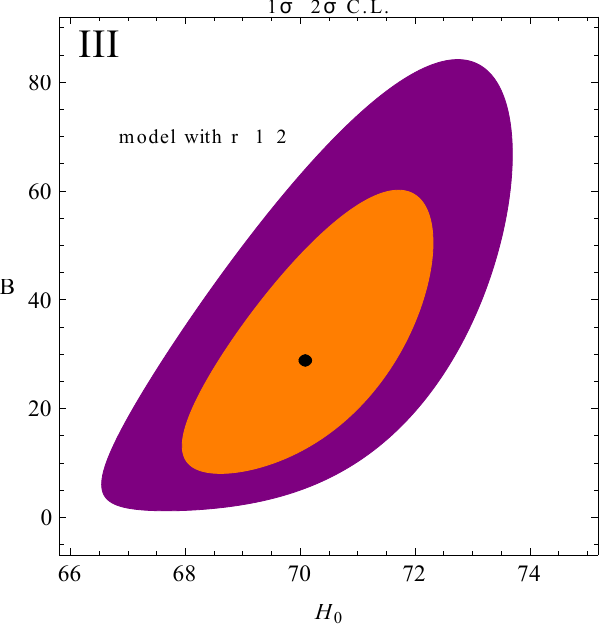}}\hskip0.06cm
\resizebox{1.8in}{!}{\includegraphics{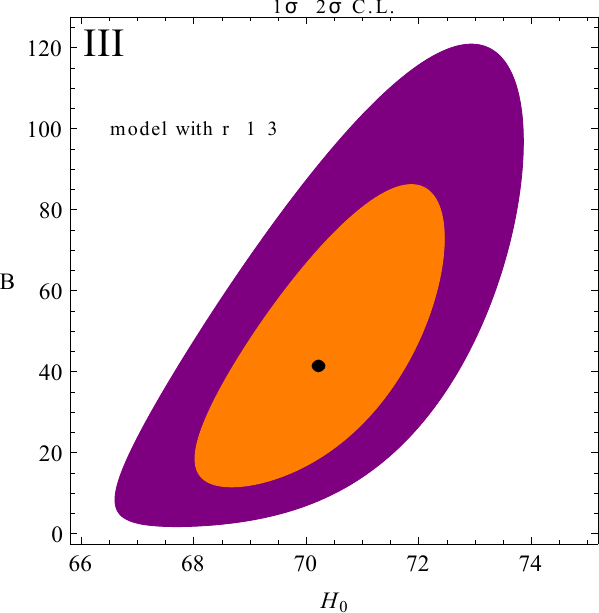}}\hskip0.06cm
\caption{\scriptsize{Two-dimensional C.L. associated with $1\sigma$, $2\sigma$ for different $\theta$ planes.}}
\label{F11}
\end{figure}

In theoretical models it is demanded that the parameters should satisfy the inequalities (i) $A>0$ and (ii) $B>0$. Some plots of the regions of $1\sigma$ and $2\sigma$ confidence levels (C.L) obtained with the standard $\chi^2$ function are shown in Fig. \ref{F11}, on the right the model with $r=1/2$ and on the left the model with $r=1/3$. The respectively estimation for the model is briefly summarized in Tables \ref{I1} and \ref{I2}. For example, some best-fitting values obtained for the parameters are, $A=49.39^{+21.69}_{-21.59}$ and $B=30.95^{+22.19}_{-17.95}$ with $\chi^{2}_{d.o.f}=0.764$ for the model with $r=1/2$ and $A=75.81^{+24.07}_{-26.84}$, $B=24.34^{+16.84}_{-12.94}$ with $\chi^{2}_{d.o.f}=0.765$ for $r=1/3$. In both cases it is satisfied the goodness condition $\chi^{2}_{d.o.f}<1$. We get the best fit at the independence parameters ($\Omega_{x}$, $\Omega_{m}$)$=$($0.714^{+0.026}_{-0.025}$, $0.249^{+0.077}_{-0.084}$) with $\chi^{2}_{d.o.f}=0.831$ for the case with $r
 =1/2$ by using the priors
($H_{0}=68$, $A=60$, $B=25$, $\ga_{m}=1.08$); therefore the present day values obtained of the dark energy and dark matter parameters are in agreement with the data released by the WMAP-9 project \cite{WMAP9} or with the data coming from the Planck Mission \cite{Planck2015XIII}. A similar result is obtained for the model with $r=1/3$ as is shown in Table \ref{I2}.

\vskip 0.1cm
\begin{table}
\centering
\scalebox{0.8}{
\begin{tabular}{|l|l|l|l|}
  \hline
  \multicolumn{4}{|c|}{${\rm 2D}$ ${\rm Confidence}$ ${\rm level}$ ${\rm for}$ ${\rm r=1/2}$} \\
  \hline
  ${\rm N^{o}}$ & ${\rm Priors}$ & ${\rm Best}$ ${\rm fits}$ & ${\rm {\chi_{d.o.f}^2}}$ \\
  \hline
  I & ($H_{0}$, $\Omega_{x}$, $\Omega_{m}$, $\ga_{m}$)=(69.2, 0.72, 0.235, 1.07) & ($A$, $B$)=($49.39^{+21.69}_{-21.59}$, $30.95^{+22.19}_{-17.95}$) & 0.764 \\
  II & ($\Omega_{x}$, $\Omega_{m}$, $B$, $\ga_{m}$)=(0.72, 0.235, 30, 1.07) & ($H_{0}$, $A$)=($70.12^{+1.49}_{-1.23}$, $59.71^{+20.81}_{-22.37}$) & 0.749 \\
  III & ($\Omega_{x}$, $\Omega_{m}$, $A$, $\ga_{m}$)=(0.721, 0.235, 60, 1.07) & ($H_{0}$, $B$)=($70.14^{+1.48}_{-1.25}$, $29.08^{+18.37}_{-16.04}$) & 0.748 \\
  IV & ($H_{0}$, $A$, $B$, $\ga_{m}$)=(68, 60, 25, 1.08) & ($\Omega_{x}$, $\Omega_{m}$)=($0.714^{+0.026}_{-0.025}$, $0.249^{+0.077}_{-0.084}$) & 0.831 \\
  \hline
  \end{tabular}}
\caption{\label{I1} \scriptsize{We show the observational bounds for the 2-D C.L. obtained in Fig. (1) by varying two cosmological parameters.}}
\end{table}

\vskip 0.1cm
\begin{table}
\centering
\scalebox{0.8}{
\begin{tabular}{|l|l|l|l|}
  \hline
  \multicolumn{4}{|c|}{${\rm 2D}$ ${\rm Confidence}$ ${\rm level}$ ${\rm for}$ ${\rm r=1/3}$} \\
  \hline
  ${\rm N^{o}}$ & ${\rm Priors}$ & ${\rm Best}$ ${\rm fits}$ & ${\rm {\chi_{d.o.f}^2}}$ \\
  \hline
  I & ($H_{0}$, $\Omega_{x}$, $\Omega_{m}$, $\ga_{m}$)=(69.2, 0.721, 0.235, 1.08) & ($A$, $B$)=($75.81^{+24.07}_{-26.84}$, $24.34^{+16.84}_{-12.94}$) & 0.765 \\
  II & ($\Omega_{x}$, $\Omega_{m}$, $B$, $\ga_{m}$)=(0.72, 0.235, 25, 1.07) & ($H_{0}$, $A$)=($70.24^{+1.43}_{-1.50}$, $80.74^{+25.96}_{-26.90}$) & 0.745 \\
  III & ($\Omega_{x}$, $\Omega_{m}$, $A$, $\ga_{m}$)=(0.721, 0.235, 60, 1.07) & ($H_{0}$, $B$)=($70.25^{+1.45}_{-1.51}$, $41.60^{+26.14}_{-22.63}$) & 0.744 \\
  IV & ($H_{0}$, $A$, $B$, $\ga_{m}$)=(68, 60, 25, 1.06) & ($\Omega_{x}$, $\Omega_{m}$)=($0.707^{+0.024}_{-0.026}$, $0.247^{+0.085}_{-0.086}$) & 0.831 \\
  \hline
  \end{tabular}}
\caption{\label{I2} \scriptsize{We show the observational bounds for the 2-D C.L. obtained in Fig. (1) by varying two cosmological parameters.}}
\end{table}

\subsection{Cosmological constraints from supernova observations}
In order to constraint the parameters of the model we use the data from a joint analysis of type Ia supernova (SN Ia) observations obtained by the SDSS-II and SNLS collaborations \cite{Betoule}. The data set includes a total of 740 spectroscopically confirmed type Ia supernovae with high quality light curves.

\begin{figure}
\centering
\resizebox{2.5in}{!}{\includegraphics{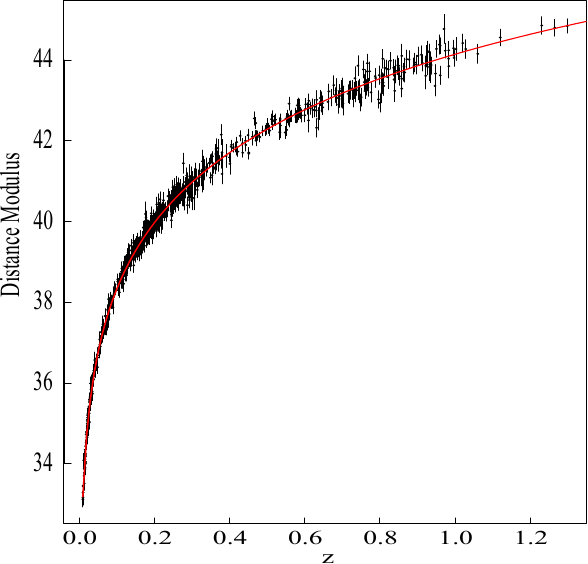}}
\caption{\label{Ajuste}\scriptsize{The distance modulus redshift relation of the best-fit model with $r=1/2$ is shown as the red line.}}
\end{figure}

For this analysis the standardized distance modulus 
\begin{equation}
\mu=5\log_{10}\left(d_{L}(z)/Mpc\right)+25,
\label{mu}
\end{equation} 
is taken into account, where the $d_L$ is the luminosity distance defined as $d_{L}(z)=c(1+z)\int_{0}^{z}\frac{dz\acute{}}{H(z\acute{})}$, $c$ is the velocity of light and $H(z)$ is the one give it in Eq. (\ref{H2}). In this case, we minimized the chi-square function defined as 
\be
\label{chimu}
\chi^2(\theta)=\sum^{N=740}_{i=1}\frac{[\mu(\theta;z_i)-\mu_{obs}(z_i)]^2}{\sigma^2(z_i)},
\ee
in order to obtain the parameters of the model.
Here $\mu_{obs}$ is the observed distance modulus used in \cite{Betoule}. The function $\mu(\theta,z_i)$ is (\ref{mu}) evaluated at $z_i$, for both models, with $r=1/2$ and $r=1/3$ respectively. In the reference \cite{Betoule} they computed a fixed fiducial value of $H_0 = 70 km s^{-1} Mpc^{-1}$, we use this value for $H_0$ to get the best-fitting values of the models. The model parameters obtained from this best-fittin analysis with supernovae observational data are showed in Table \ref{Sn}. The Hubble diagram for the JLA sample and the model fit are shown in Fig. \ref{Ajuste}.

\begin{figure}
\centering
\resizebox{1.8in}{!}{\includegraphics{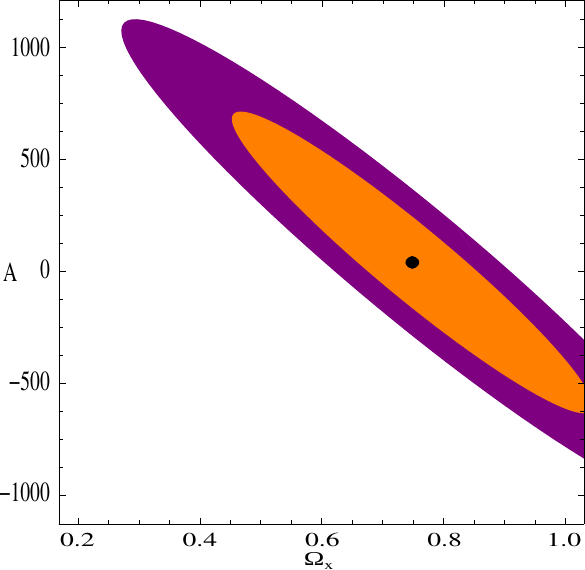}}\hskip0.06cm
\resizebox{1.8in}{!}{\includegraphics{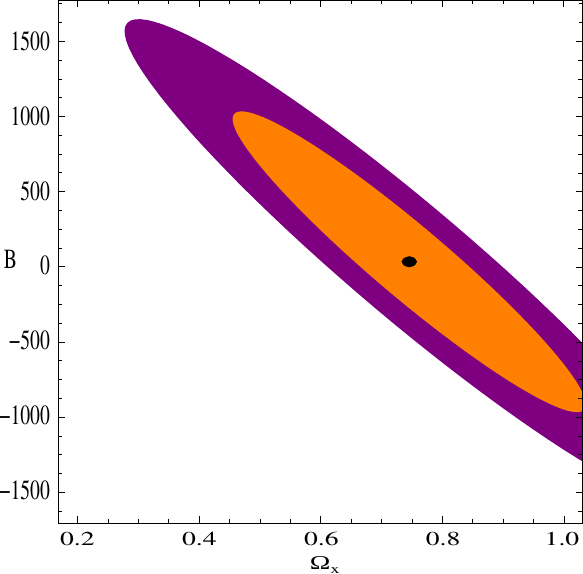}}\hskip0.06cm
\caption{\scriptsize{Two-dimensional C.L. associated with $1\sigma$, $2\sigma$ for different $\theta$ planes.}}
\label{FSn}
\end{figure}

The best fit value for the dark energy parameter is $\Omega_x=0.745\pm0.032$ for the model with $r=1/2$ and $\Omega_x=0.746\pm0.033$ for $r=1/3$. These values are in agreement with the ones founded with the Hubble data in the last section and with the observations \cite{Planck2015XIII}, \cite{Planck2015XIV} \cite{WMAP9}. We also plot the confidence contours of $(68.3\%)$ and $(95.4\%)$ for $\Omega_x$ and the parameters $A$ and $B$ in Fig. \ref{FSn}.

Moreover, the best fit values for the dark matter parameter and the barotropic matter index are $\Omega_m=0.245\pm0.123$ and $\gamma_m=1.039\pm0.188$ for the model with $r=1/2$ respectively and $\Omega_m=0.224\pm0.131$ and $\gamma_m=1.082\pm0.177$ with $r=1/3$.

\begin{table}
\centering
\scalebox{0.95}{
\begin{tabular}{|l|l|l|l|}
  \hline
  ${\rm Model}$ & \qquad $ A$ & \qquad $B$ & ${\rm {\chi_{d.o.f}^2}}$ \\
  \hline
  ${\rm r=1/2}$ & $37.59\pm 36.21$ & $31.25\pm 56.04$ & 0.863 \\
  \hline
  ${\rm r=1/3}$ & $58.38\pm 57.59$ & $11.49\pm 24.53$ & 0.859 \\
  \hline
 \end{tabular}}
 \caption{\label{Sn} \scriptsize{The best-fitting values obtained using the data base JLA.}}
\end{table}

\section{Other relevant parameters} 
\begin{figure}
\centering
\resizebox{3.3in}{!}{\includegraphics{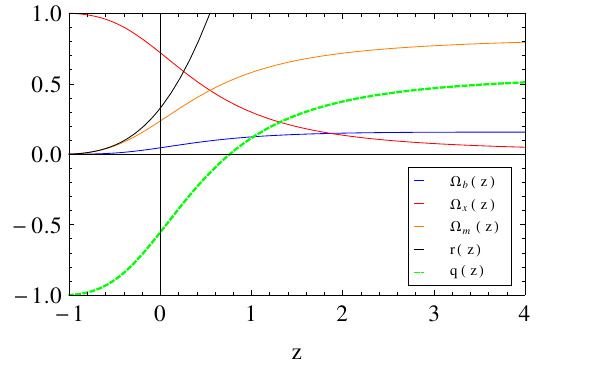}}
\caption{\label{T-12}\scriptsize{Plot of $\Omega_{b}(z)$, $\Omega_{x}(z)$, $\Omega_{m}(z)$, $r(z)$, and $q(z)$, using the best-fit values obtained with the Hubble data for different $\theta$ planes, for the model with $r=1/2$.}}
\end{figure}

For the models with $r=1/2$ and $r=1/3$ the behavior of the density parameters $\Omega_x$, $\Omega_m$, and $\Omega_b$ nearly close to $z=0$ is described in Fig. \ref{T-12}. As we well know, the dark energy is in particular the main source responsible of the Universe acceleration; far away from $z=1$ the Universe is dominated by the dark matter which it is responsible of the structure formation. Note that these models are asymptotically de Sitter when $z\rightarrow-1$ and the total energy density tends to a constant value. 

Other cosmological relevant parameter is the deceleration parameter at the present time $q(z=0)=q_0$. The Figure \ref{T-12} shows the behaviour of the deceleration parameter with the redshift. In particular, the present-day value of $q(z=0)$ is between $[-0.56; -0.54]$ as can be seen from Table \ref{II2}.

\begin{table}
\centering
\hspace{-0.5cm}
\begin{minipage}[b]{0.38\linewidth}
\centering
\scalebox{0.75}{
\begin{tabular}{|l|l|l|l|}
  \hline
  \multicolumn{4}{|c|}{${\rm Cosmological}$ ${\rm parameters}$ ${\rm for}$ ${\rm r=1/2}$} \\
  \hline
  ${\rm N^{o}}$ & $q(z=0)$ & $\Omega_{x}(z\approx1100)$ & $\Omega_{x}(z\approx10^{10})$ \\
  \hline
  I & -0.56 & 0.0032 & 0.00012 \\
  \hline
  II & -0.56 & 0.0032 & 0.00012 \\
  \hline
  III & -0.56 & 0.0031 & 0.00012 \\
  \hline
  IV & -0.54 & 0.0020 & 0.00004 \\
  \hline
 \end{tabular}}
\end{minipage}
    \hspace{1.3cm}
    \begin{minipage}[b]{0.38\linewidth}
    \centering
\scalebox{0.75}{
\begin{tabular}{|l|l|l|l|}
  \hline
  \multicolumn{4}{|c|}{${\rm Cosmological}$ ${\rm parameters}$ ${\rm for}$ ${\rm r=1/3}$} \\
  \hline
  ${\rm N^{o}}$ & $q(z=0)$ & $\Omega_{x}(z\approx1100)$ & $\Omega_{x}(z\approx10^{10})$ \\
  \hline
  I & -0.55 & $2.7\times10^{-6}$ & $6.3\times10^{-15}$ \\
  \hline
  II & -0.56 & $3.3\times10^{-6}$ & $1.3\times10^{-14}$ \\
  \hline
  III & -0.56 & $5.4\times10^{-6}$ & $2.1\times10^{-13}$ \\
  \hline
  IV & -0.54 & $3.7\times10^{-6}$ & $2.3\times10^{-14}$ \\
  \hline
 \end{tabular}}
 \end{minipage}
 \caption{\label{II2} \scriptsize{We show the cosmological parameters derived from the best fits value of 2-D C.L. obtained in Tables (\ref{I1}) and (\ref{I2}) by varying two cosmological parameters.}}
\end{table}

We also determined the variation of the dark energy parameter behind recombination or big-bang nucleosynthesis epochs \cite{Luis7}, \cite{Sanchez} and compared with the severe bound for each epoch. This can be considered as a complementary tool for testing our models. One of the last constraints on early dark energy (ede) comes from the Planck TT, TE, EE+lowP+BSH data: $\Omega_{ede} < 0.0036$  at $95\%$ C.L \cite{Planck2015XIV}. We  found that $\Omega_{x}(z\simeq 10^{3})$ is over the interval $[0.0020,0.0032]$ for the model with $r=1/2$ and $[2.7\times 10^{-6},5.4\times 10^{-6}]$ for $r=1/3$, so our estimations satisfied the bound reported by the Planck mission [see Tables \ref{II2}]. In regard to the bound reported from the joint analysis based on  Euclid+CMBPol data, $\Omega_{ede} <0.00092$ \cite{HollEuc}, \cite{Cala1}, the model with $r=1/2$ does not satisfy the severe bound, but the model with $r=1/3$ fulfil the bound reported. Around $z=10^{10}$, in the nucleosynthesis epoch, we have $\Omega_{x}$ between $[10^{-15}; 10^{-13}]$ at the $1\sigma$ level, therefore the model with $r=1/3$ is in concordance with the conventional BBN processes that occurred at a temperature of $1 {\rm Mev}$ \cite{WrightBBN}. For the best fit values obtained using the JLA sample for the models, the dark energy behind recombination is $\Omega_{x}(z\simeq 10^{3})=0.0056$ and $\Omega_{x}(z\simeq 10^{3})=1.05\cdot10^{-6}$ with $r=1/2$ and with $r=1/3$ respectively. For the nucleosynthesis epoch, the values are $\Omega_{x}=0.001$ for the model with $r=1/2$ and $\Omega_{x}=2.18\cdot10^{-15}$ with $r=1/3$. These values coincide with the ones obtained by constraining with the Hubble data.

\section{Discussions}
In the present letter a Universe that presents a particular interaction in the dark sector has been analyzed. It was found that, when the interaction depends on the scale factor as $Q=-\sqrt{B}a^{-3r}(2A+\sqrt{B}a^{-3r})$, and when the dark sector is characterized by a barotropic equation of state $p=(\gamma-1)\rho$, then the flat Emergent Universe solution that was already presented in \cite{Muk} appears. However, it should be emphasized that these Emergent Universe solutions were obtained in these references by using a non-linear equation of state, which is a particular case of the Chaplying gas, and not with the interactions considered by us. So our result can be considered as original.
The reason for which we analyzed this particular interaction and no others is because we have found that this can cast the solution of an Emergent Universe \cite{Paul1}, \cite{Ghose}.  Interactions with  $\gamma_x\neq0$, together with different interactions or models with variable $\Lambda$ such as those studied in \cite{Sola4}, \cite{Sola5} will be considered in a separate work.

The comparison with observational data was carried out by considering the parameter values $r=1/2$ and $r=1/3$, and the remaining cosmic set of parameters has been constrained by using the updated Hubble data, the JLA supernova data and the severe bounds for dark energy found at early times. We have shown that both models interpolate between a cold dark matter regime and a De Sitter phase in the asymptotic future. 

On the observational side, the best-fit values at 2$\sigma$ level, using the Hubble data, for the parameters of the model are represented in Fig. \ref{F11} and Tables \ref{I1} and \ref{I2}. We observe that the obtained constant values of the models are $A=49.39^{+21.69}_{-21.59}$ and $B=30.95^{+22.19}_{-17.95}$ with $\chi^{2}_{d.o.f}=0.764$ for the model with $r=1/2$ and $A=75.81^{+24.07}_{-26.84}$, $B=24.34^{+16.84}_{-12.94}$ with $\chi^{2}_{d.o.f}=0.765$ for $r=1/3$, where $A>0$ and $B>0$ for both cases. They satisfy the goodness condition $\chi^{2}_{d.o.f}\approx1$. The best fit is obtained at the independence parameters ($\Omega_{x}$, $\Omega_{m}$)$=$($0.714^{+0.026}_{-0.025}$, $0.249^{+0.077}_{-0.084}$) with $\chi^{2}_{d.o.f}=0.831$ by using the priors ($H_{0}=68$, $A=60$, $B=25$, $\ga_{m}=1.08$) for the case with $r=1/2$. The values obtained for the dark energy and dark matter density parameters are in agreement with the data coming form the WMAP-9 project \cite{WMAP9} or with the data realised by the Planck Mission \cite{Planck2015XIII}, see Table \ref{I1}. For the model with $r=1/3$ as it is shown in Table \ref{I2}, we get a similar result. 
In the same line, the best-fit values using the JLA sample of the parameters of the models are $A=37.59\pm36.51$ and $B=31.25\pm56.04$ with $\chi^{2}_{d.o.f}=0.863$ for the model with $r=1/2$ and $A=58.38\pm57.59$ and $B=11.49\pm24.53$ with $\chi^{2}_{d.o.f}=0.859$ for $r=1/3$. The values obtained of the dark energy and dark matter density parameter as we see in section 3.2 are in agreement with the data.
In addition, the amount of early dark energy has been estimated, i.e. the energy density parameter in the radiation era. We found that the two models fulfil the severe bounds of $\Omega_{x}(z\simeq 1100)<0.009$ at the $2\sigma$ level of Planck. But the model with $r=1/2$ did not satisfy the severe bound reported by the joint analysis based on  Euclid+CMBPol data, $\Omega_{ede} <0.00092$ \cite{HollEuc}, \cite{Cala1}, while the model with $r=1/3$ does it.


The central aim of the work is to show that a linear equation of state and the proposed interaction $Q$ in the dark sector, instead of a mechanism that makes each of them more complex, recover the solution of Emergent Universe models. 
We recognize the limitation of the model but it does not remove the fact that it is an original work and it deserves to be studied. In fact, to show that an interaction can lead to the same kind of universe as a non-barotropic state equation is something that deserves to be investigated and the possibility of generalizing in a future work is not ruled out. Under this assumption, we leave for a future research the consideration of the BAO scale, the CMB dates and the growth of perturbation. Nevertheless the analysis performed here over the updated observational Hubble data and the JLA supernovae data, which predict dark densities close to the observations, is enough to prove the viability of the approach proposed by us. These results should be considered in future investigations and discussions.

\section*{Acknowledgements}
The authors are supported by CONICET. 

\vskip 1cm


\end{document}